\documentclass[
superscriptaddress,
twocolumn,
showpacs,
showkeys,
pra,
floatfix,
]{revtex4-1}

\usepackage{amssymb}
\usepackage{graphicx}
\usepackage{bm,bbm}
\usepackage{color}

\newcommand{\One}{\mathbb{I}}

\newcommand{\al}{\textit{et al.}}
\newcommand{\eg}{\textit{eg.}}
\newcommand{\mt}{\mathrm }
\newcommand{\tauk}{\tau_{\mathrm{k}}}

\newcommand{\rmi}{\mathrm{i}}
\newcommand{\rmd}{\mathrm{d}}
\newcommand{\rme}{\mathrm{e}}

\newcommand{\h}{\hat}
\newcommand{\p}{^}

\newcommand{\la}{\langle}
\newcommand{\ra}{\rangle}

\newcommand{\Hs}{H_{\mathrm{s}}}
\newcommand{\Hrf}{H_{\mathrm{rf}}}

\newcommand{\Henv}{H_{\mathrm{env}}}
\newcommand{\Hint}{H_{\mathrm{int}}}

\begin{document}
%\title{Non equilibrium stationary states of a kicked linear chain of spins coupled to a finite temperature bath}
\title{Non equilibrium stationary states of a dissipative kicked linear chain of spins}

\author{P. C. L\' opez V\' azquez}
\affiliation{Departamento de Ciencias Naturales y Exactas, Universidad de Guadalajara,  
Carretera Guadalajara - Ameca Km. 45.5 C.P. 46600. Ameca, Jalisco, M\'exico.}

\author{A. Garc\'ia}
\affiliation{Departamento de Matem\'aticas, Universidad de Guadalajara,
  Blvd. Marcelino Garc\'\i a Barragan y Calzada Ol\'\i mpica, 
  C.P. 44840, Guadalajara, Jalisco, M\' exico.}
  
\author{G. S\'anchez-Gonz\'alez}
\affiliation{Centro de Investigaci\'on Sobre Enfermedades Infecciosas, Instituto Nacional de Salud P\'ublica, Universidad No. 655,
 C.P. 62100, Cuernavaca, Morelos. M\'exico.}

\begin{abstract}
We consider a linear chain made of spins of one half in contact with a dissipative 
environment for which periodic delta-kicks are applied to the qubits of the linear chain in two different
configurations: kicks applied to a single qubit and simultaneous kicks applied to two qubits 
of the linear chain. In both cases the system reaches a non-equilibrium stationary condition
in the long time limit. We study the transient to the quasi stationary states and
their properties as function of the kick parameters in the single kicked qubit case and report 
the emergence of stationary entanglement between the kicked qubits when simultaneous kicks are applied. 
For doing our study we have derived an approximation to a master equation which serves us to analyze the
effects of a finite temperature and the zero temperature environment.
\end{abstract}

\maketitle

\section{\label{I}Introduction}
The understanding of the creation of the stationary states on
open systems which are subject to driving forces placing them out
of equilibrium is of great importance in the field of complex
systems and comparable in importance to the fundamental ideas of the
stationary states in physical statistics.
One of the most important contributions in this field
was given by Haken in its theoretical description 
of the laser dynamics~\cite{Haken1970}.
His results brought some first insights on emergent properties appearing in complex systems 
due to a cooperative behavior of driving and dissipative forces 
acting on them. These ideas became the fundamental principals of the 
theory of synergetics created by Haken himself~\cite{Haken2004}.
Open quantum chaotic systems are systems subject to these two type of mechanisms. 
Although quantum chaotic systems have been firstly studied 
in the context of environmental systems~\cite{LutWei99,GS02b,CaGoSe2014},
their interaction with other degrees of freedom acting as a finite temperature
reservoir is inevitable.
In this sense, Gorin~\al~\cite{MoGoSe2015} have studied the dynamics of a qubit in contact to a near chaotic 
environment based on random matrix ensemble which in turn 
is coupled to a heat bath, considered as a far 
environment affecting the qubit through the chaotic environment.
For this tripartite type of system, they have found a recovery in 
the purity of the qubit when the coupling of the chaotic environment
to the heat bath was increased. This is a counterintuitive effect that
may be related to the cooperative mechanisms of dissipation and 
driving forces giving rise to emergent properties in the near environment
that decouples the interaction of the qubit with the near environment.
Additionally, there has been some recent developments
concerning the thermodynamical properties of non-equilibrium quantum 
systems~\cite{LanHol14,KetWus10} and in particularly 
for quantum kicked systems in contact with a
thermal reservoir~\cite{PrLoGo2017,LoGa2016} where the non-equilibrium dynamics
are introduced with the help of time-dependent periodic delta-kicked
potentials~\cite{GCZ97,BilGar09,DaDo2005}. 
In this context, the freedom to choose strong or weak
interactions with the kicks and with the environment,
have open up new interesting features on the  
the emergent thermodynamical properties of the 
non-equilibrium stationary states or "quasi stationary" states
reached by the system in the long time limit.
This quasi stationary condition is reached when 
the system asymptotically gets rid of its dependence
on the initial conditions and enters into a limit cycle dynamics
in which the amount of energy received by a single kick 
equals the amount of energy dissipated into the environment
between two consecutive kicks. At this regime, the observables are
obtained by averaging the desired quantities over
the fluctuations that appear in the system as a consequence of the kicks 
and the features related to the the quantum kicked systems like resonances
and anti-resonances~\cite{GCZ97,BilGar09,DaDo2005,El2008} or
localization, consequence of the kicks~\cite{Fi1982,An1978,Ad2009}, they disappear 
and only the strength of the kicks and the period of the kicks 
become the relevant quantities in the formation of the quasi-steady states.\\
In this paper we want to report our studies of the 
formation of a quasi stationary states in a liner chain made of nuclear spins which has been
a model of certain types of quantum computer devices based on a chain of 
nuclear paramagnetic atoms~\cite{Be2000,GV2003}. 
Contrary to the kicked harmonic oscillator or the
kicked rotator~\cite{PrLoGo2017,LoGa2016} where one can indefinitely populate 
states by the application of kicks since these systems
possess an unbounded spectrum; the linear chain is a finite 
dimensional system whose dimension, (dim$\mathcal{H}=2^N$), depends on the number of qubits
$N$, and one cannot indefinitely populate states by the application of repeated kicks.
Therefore the formation of the quasi-steady states has to be in general a quite different situation.\\
It is worth to mention that this model has certain similitudes 
on what has been described in~\cite{Vi1998,Re2009,Uh2008,Mo2006,Liu2013} 
regarding the dynamical decoupling effects 
of a kicked qubit when the kics are done very fast compared to the characteristic times of evolution 
of the system. However this is not our case in the sense that we will be dealing 
with a finite number of qubits in a chain for which at most a pair of them will only 
be subject to the kicks. For having dynamical suppression in this model one would have to be able to kick 
very fast, each one of the qubits of the linear chain. 
The aim of this paper is to present the properties of the quasi stationary states reached by the system
under different configurations of the kicks and to 
convince the reader that it is possible to produce exotic forms of steadiness 
such as entanglement between qubits of the linear chain.\\
This paper is organized as follows: In section \ref{Mod} we describe the model of
the linear chain subject to kicks and in contact to a thermal bath. 
We also  present an approximation of a master equation for the model of
the linear chain in contact  with the thermal bath
and establish certain parameters of the system we will be using along the paper.
In section~\ref{singkicks} we show the transient dynamics and some properties 
of the quasi stationary states when the kicks are applied to single qubits 
of the linear chain when the finite temperature and zero temperature limits 
are considered in the interaction with the bath.
In section~\ref{simkicks} 
we study the situation where simultaneous kicks are applied to a
couple of qubits of the linear chain and focus on the formation 
of stationary entanglement between the pair of kicked qubits. 
Finally in section~\ref{sum} we give a summary of our results and at 
the appendix~\ref{app} we present our derivation of the master equation
for this model.

\section{\label{Mod}The model}
The model consist on linear chain made of $N$ spins of one half 
or qubits, interacting with a non-homogeneous stationary magnetic field 
directed along the $z$-axis. The linear chain lies in an
angle of $\cos\theta = 1/\sqrt{3}$ with respect to the $z$-axis
in order to eliminate the dipole-dipole interaction between
the qubits and only Ising type of interaction in
the $z$ component to second neighbors is assumed. 
The Hamiltonian of the ideal insulated linear chain is given by:
\begin{equation}\label{hs}
 \Hs = -\sum_{l=1}^{N}\omega_l s_l^z - {J\over \hbar}\sum_{l=1}^{N-1}s_l^zs_{l+1}^z
 -{J'\over \hbar}\sum_{l=1}^{N-2}s_l^zs_{l+2}^z.
\end{equation}
with $\omega_l$ being the Larmor frequencies of each one of the $N$ qubits in 
the linear chain and  $J$ and $J'$ quantify the coupling strength to the
first and second neighboring qubits respectively. 
This system is based on a quantum computer model of 
a linear chain of nuclear paramagnetic atoms 
interacting with a RF-field 
which is able to perform Rabi transitions between 
the states of the linear chain when 
the proper angular frequency of the RF-field is chosen
~\cite{Be2000,GV2003,LoLo2012a,LoLo2012b}.
In this paper we will assume that the RF part of the field is
switched off and only the $z$-component of the magnetic field,
which generates a precession movement of the magnetic moments 
of the nuclear atoms will be considered. 
The eigenbasis of the Hamiltonian 
$\Hs$ is named as $\{|\alpha_{N}\dots\alpha_{1}\rangle\}$ 
for $\alpha_j=0,1$ with $j$ labeling the $j$-th spin in the linear chain. 
The action of the $j$-th spin operators in this basis are
defined as: $s_j^z|\alpha_j\rangle= 
{\hbar\over 2}(-1)^{\alpha_j}|\alpha_j\rangle$, $s_j^+|\alpha_k\rangle= 
\hbar \delta_{\alpha_j,0}|1\rangle$, and 
$s_j^{-}|\alpha_j\rangle=\hbar\delta_{\alpha_j,1}|0\rangle$. 
The elements of this basis forms a register of 
$N$-qubits with a total number of $2^N$ registers, which is the dimensionality
for the Hilbert space.\\
In our model, the interaction with the environment plays a crucial role. 
For that reason, we assume that the linear chain is
immerse in a dissipative finite 
temperature thermal environment consisting on
a quantized radiation field with an infinite number
of radiation modes~\cite{BrePet02,LoLo2012b}.
The Hamiltonian of the bath can be described in general terms as
a large set of harmonic oscillators with the vacuum energy shifted out. 
The interaction Hamiltonian is described through 
the dipole approximation:
\begin{equation}\label{hint}
 \Hint=\sum_{i,l}^{\infty,N} g_{il} s_l^{+}a_{i} 
 + g_{ij}\p{*}s_l^{-}a^{\dag}_{i}.
\end{equation}
This type of interaction accounts for exitation-de exitation
processes in the system through the coupling to the bath of oscillators having
characteristic frequencies near the resonant frequencies of the linear chain.
The $g_{ij}$ are the coupling strengths of the spins to
the thermal bath and $a_i(a^{\dag}_i)$ are the rising (lowering) operators in the number
of photons in the bath. For this model of interaction we have derived
a master equation following the weak coupling approximation and the 
Born-Markov limit~\cite{BrePet02}.
The details of this derivation are described in the appendix \ref{app}
for which the RF part of the magnetic field has also been included.
An important remark about the model of dissipation is that the super operator
in the master equation that describes 
the non unitary evolution of the system does not has a Lindblad form,
nevertheless it describes properly rates of dissipation for
the different non-equidistant energy levels of $\Hs$.\\
Finally, the system subject to periodic kicks that drives the
system out of equilibrium. These kicks represents a series of
rotations of the qubit or qubits around a certain axis.
They can be understood as successive unitary 
transformations in the wave function happening at fixed
intervals of time $t_{\mt k}$ produced by an additional external microwave 
field~\cite{Uh2008}.
Additionally, no coupling to the bath is assumed
during their application since it is assumed that 
the kick produces instantaneous changes in the system,
see \eg~\cite{PaFiKaUh2008} for the 
application of pulses with a finite duration.
We use a periodic delta-kicked potential to describe the action of the pulses
done to the jth-qubit of the linear chain:
\begin{equation}\label{hk}
  V_j( t ) = {\kappa}s^{\eta}_j
  \sum_{n=-\infty}^{\infty}\delta(t - n t_{\mt k})\; . 
\end{equation}
Here $\kappa$  represents the angle of rotation of the qubit 
about the $\eta$-axis ($\eta=x,y$ or $z$). The subindex $j$ labels the spin in the linear
chain subject to the kicks and $t_{\mt k}$ is the 
period of the kicks which we kept fixed in our simulations.

\subsection{\label{diml}Dimensionless model and implementation of the dynamics}
The dynamics of the system are described in terms
of two alternating autonomous quantum maps. 
One map describes the dissipative non-unitary 
dynamics under a master equation which we present hereafter,
and the second map is the unitary
transformation produced by the kicks. 
We will use a dimensionless description of the dynamics through the 
Pauli matrices representation of spins: 
$\vec{\sigma} =2\vec{s}/\hbar$ 
for each of the spins in the linear chain. 
Also we measure everything in terms of a
dimensionless time scale by choosing the largest 
Larmor frequency of the qubits in 
the linear chain and measure everything 
in terms of the period of precession of this spin. 
We define our dimensionless time as: $\tau = \omega_A t$, for
$\omega_A = \max_{j=1,...N}\omega_j$. 
With these redefinitions 
we write the master equation of the system as:
\begin{equation}\label{medl}
\rmi {\rmd \over \rmd \tau }\varrho_{\mt s} = 
[\Hs , \varrho_{\mt s}] + \rmi \mathcal{D} [\varrho_{\mt s}]
\end{equation}
where $\Hs$ takes the form: 
\begin{eqnarray}\label{hsc1}
 \Hs &=& - {1\over 2}\sigma_A^z-{1\over 2}\sum_{j=1}^{N-1}\delta_j \sigma_j^z \\\nonumber
  &&- {\chi\over 4} \sum_{j=1}^{N-1} \sigma_j^z\sigma_{j+1}^z - {\chi'\over 4} \sum_{j=1}^{N-2} \sigma_j^z\sigma_{j+2}^z  
\end{eqnarray}
with $\delta_{j}= \omega_{j}/\omega_A < 1$, $\chi = J /\omega_A$ and $\chi'= J'/\omega_A$, 
and the term that accounts for the dissipative behavior due to the interaction with the thermal bath has
the form (see \ref{app}):
\begin{eqnarray}\nonumber
\mathcal{D}[\varrho_{\mt s}] &=&
- \sum_{l=1}^{N} \beta_l\left\{\left[ \h{\mt{O}}\p{(1)}_{l}\,\sigma_l\p{+},\sigma_l\p{-}\varrho_{\mt s}\right]
+\left[ \varrho_{\mt s}\sigma_l\p{+},\sigma_l\p{-}\,\h{\mt{O}}\p{(1)}_{l}\right]\right.\\\nonumber
&&+\left.\left[ \h{\mt{O}}\p{(2)}_{l}\,\sigma_l\p{-},\sigma_l\p{+}\varrho_{\mt s}\right]
+\left[ \varrho_{\mt s}\sigma_l\p{-},\sigma_l\p{+}\,\h{\mt{O}}\p{(2)}_{l}\right]\right\}\\\label{disc1dl}
\end{eqnarray}
where $\beta_l=\gamma_l/4\omega_A$ with $\gamma_l$ being a parameter that accounts for
the strength of coupling to the environment
and $\h{\mt{O}}^{(1,2)}$ is an operator that depends on the dimensionless temperature
of the bath $ D = k_B T/\omega_A \hbar$, (see  \ref{app}).
The  zero temperature limit is assumed when the dimensionless temperature of the
bath is sufficiently small compared to the dimensionless transition energies 
of the linear chain. In this limit, one makes $D\rightarrow 0$, 
and the temperature dependent operators on the super operator (\ref{disc1dl}) become:
$\h{\mt{O}}\p{(1)}_{l}\!(D\rightarrow 0)\rightarrow \h{\Omega}^3_l$ 
and $\h{\mt{O}}\p{(2)}_{l}\!(D\rightarrow 0)\rightarrow 0$.
In this limit, the dissipative term of the master equation describes a
pure spontaneous emission process.\\
The application of the kicks can be regarded as instantaneous changes of the wave
function. The kicks are done after the system has evolved a certain period of time 
$\tauk=\omega_A t_{\mt {k}}$, only in contact with the heat bath.
The unitary transformation representing the kick to the $j$th qubit can be described by the unitary operator: 
\begin{eqnarray}\nonumber
R_{\eta,j}^{\kappa} &=& \rme^{-\rmi \kappa \sigma^{\eta}_j/2}\\\label{rkick}
&=&\cos{\kappa/2}+\rmi\sigma^{\eta}_j\sin{\kappa/2}
\end{eqnarray}
such that, if $\varrho_{\mt s}(\tauk)$ is the solution of the master
equation (\ref{medl}), then the application of the kick will be represented
by the unitary transformation of the system: 
$R_{\eta,j}^{\kappa}\varrho_{\mt s}(t_{\mt k})R^{\kappa\dag}_{\eta,j}$.
After the application of the first kick, a new configuration
in the states of the system will appear and afterwards, the system will evolve
again non-unitarily in contact with the heat bath alone
until the next kick happens at a new equally distant interval
of time $\tauk$,  ($\tau=2\tauk$), and this process repeats several times
until the system reaches a quasi stationary condition.
In the following, we will set the number of qubits of the linear 
chain to 3, ($N=3$) since is less expensive in time computer consuming and 
the generalities of our results could easily been extrapolated to a larger 
number of qubits. The dimension of the Hilbert space is $2\p{3} = 8$
and we label the qubits as A, B and C. The states of the linear chain
form a register defined by $|\mt{ABC}\rangle$ with $\mt{A,B,C}=0,1$, 
and we use a decimal notation to represent to the different states of the
system: $|1\rangle=|000\rangle$, $ |2\rangle=|001\rangle$, $|3\rangle=|010\rangle$,
$|4\rangle=|011\rangle$, $|5\rangle=|100\rangle$,
$|6\rangle=|101\rangle$,  $|7\rangle=|110\rangle$ and $|8\rangle=|111\rangle$. 
We will also assume that the three different qubits are equally coupled to the thermal bath  
at a definite value $\beta$.

\subsection{\label{par}Parameters}
In dimensionless units as described above, we set the following 
values for the Larmor frequencies and Ising interaction constants:
$\delta_A=1$, $\delta_B=0.5$, $\delta_C=0.25$, $\chi=0.15$ and $\chi'=0.1$
because for these values the system has a non-degenerate spectrum which makes 
easier to analyze the results.
The values of $\kappa$ represent the angle of rotation  and they must
lie between 0 and $2\pi$. The later represents a full rotation of the qubit
and for this angle and for $\kappa=0$, the kicks have no effect on the linear chain.
We will use different angles of rotation and directions of rotations through the paper.
In the dimensionless description, the choices we do for the period of the kicks 
are $\tauk = 4\pi/q$ where $q$ is a positive number different from zero 
which will be varied to obtain different results. 
These choices sets the period of the kicks to be commensurable to the period of qubit A.
Finally we set the parameters of the bath to $\beta=0.1$ and the dimensionless temperature
parameter to $D=1$ for the finite temperature limit, and $D=0$ for the zero temperature limit.
The initial condition we use in our simulations is the the excited state of the linear chain:
$|\psi\rangle= |111\rangle=|8\rangle$.

\section{\label{singkicks}Transient dynamics and
quasi-steady states of sigled kicked qubits}
\begin{figure}[!htbp]
  \begin{center}
        \resizebox{85mm}{!}{\includegraphics{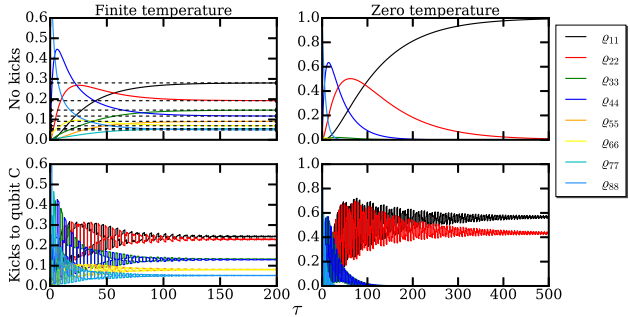}} 
          \caption{\label{fig1}
          The figures shows the diagonal elements of the density matrix for:
          $\delta_B=0.5$, $\delta_C=0.25$, $\chi=0.15$, $\chi'=0.1$ 
          for the finite temperature limit ($D=1$) and the zero temerpature limit 
          ($D=0$) and  coupling strength to the bath $\beta=0.1$. The first row shows 
          the case when no kicks are applied and the second row shows the case when 
          periodic kicks are applied to qubit C with a kick strength of $\kappa=\pi/2$ and a
          period of the kicks of $\tauk=\pi/2$.} 
  \end{center}
\end{figure}

We begin by showing comparison of the transient dynamics
between the diagonal elements of the density matrix without kicks to the dynamics 
with periodic kicks with period $\tauk=\pi/2$ applied to qubit C.
This is shown in figure \ref{fig1}.
When no kicks are applied, the finite temperature limit yield stationary
states corresponding to a Gibbs distribution (dashed black lines), and for the
zero temperature limit the system reaches the ground state as
a spontaneous emission process takes place. 
When kicks are applied to a single qubit (second row of figure \ref{fig1}),
the system reaches a quasi stationary condition which is characterized
by fluctuations around a certain averaged value. These fluctuations are seen
in the figure as discontinuities happening at the moment when a kick is done.

\begin{figure}[!htbp]
  \begin{center}
      \resizebox{85mm}{!}{\includegraphics{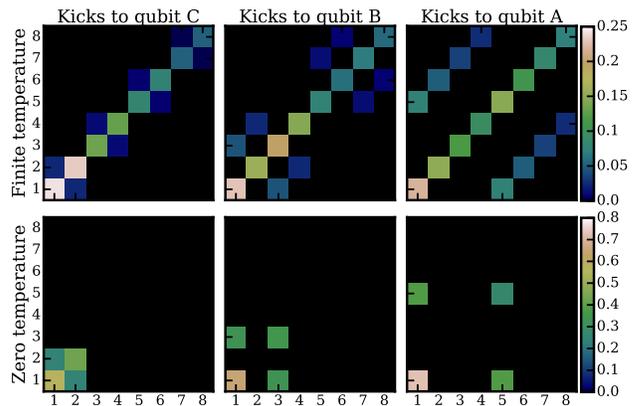}} 
          \caption{\label{fig2}
          The figure shows the matrix elements of the matrix density of the linear chain
          for periodic kicks to qubit C B and A at the quasi stationary regime for 
          kick strengths of: $\kappa=\pi/2$ and period of the kicks: $\tauk=\pi/2$,
          and  $\delta_B=0.5$,  $\delta_C=0.25$, $\chi=0.15$, $\chi'=0.1$. The coupling
          strength to the bath  is: $\beta=0.1$ and $D=1$ for the finite temperature case
          and $D=0$ for the zero temperature limit.}
  \end{center}
\end{figure}

The joint action of the bath and kicks generate stationary states 
that posses a certain degree of superposition as one can notice
in the shortened distance between the diagonal elements
associated to the transitions of the kicked qubit, \eg~at the 
finite temperature limit and according to 
the quantum register defined as $|ABC\rangle$,
the states $|AB0\rangle $ lie closer to the state $|AB1\rangle$,
for $A,B=0,1$. This superposition is more noticeable at 
the zero temperature limit, (bottom left sub figure in \ref{fig1}),
since now the effect of the bath is to drive the system to the ground state
while the kicks pulls up 
the state corresponding to the superposition while the ground state is dragged down.
In figure~\ref{fig2}, the density matrices at the quasi stationary regime 
are plotted for the zero temperature limit and the finite temperature limit
and when the kicks are done to the three different qubits. 
In this figure one notices that the coherent terms correspondent to the superposition 
of states of the kicked qubit have a non-zero value
regardless the inherent decoherence induced by the bath.
The superposition appearing in the system is in fact resilient
to the environment as they are the result of both mechanism of dissipation and kicks acting 
together over the linear chain \eg~when kicks are done to qubit A at the zero temperature limit
(bottom right sub figure in figure \ref{fig2}), there is 
superposition between the state $|1\rangle=|000\rangle$ and the 
state $|5\rangle=|100\rangle$ which appears as a consequence
of the bath attempting to  drive the system to the ground state $|1\rangle=|000\rangle$ 
making it the most likely state while the action of repeated kicks to qubit A
are always creating superpositions of states of qubit A thus, at the quasi 
stationary regime, the kicks are only acting on the ground state creating a superposition between 
this one and the state $|5\rangle=|100\rangle$ which is the state that corresponds to the single transition 
of qubit A. This explanation describes the resultant quasi steady states reached by the system when extrapolated 
to the cases when the other qubits are kicked and to the finite temperature limit where now 
the bath drives the system into a mixture of states (Gibbs distribution). We will discuss more about 
the super position states later on. \\ 

\begin{figure}[!htbp]
  \begin{center}
      \resizebox{85mm}{!}{\includegraphics{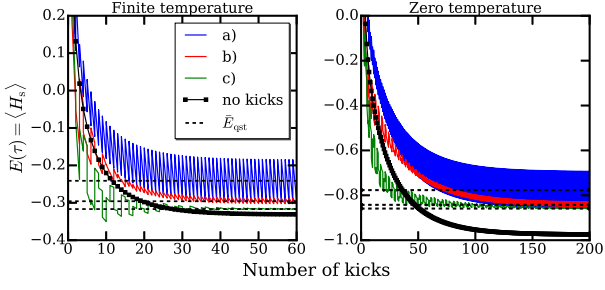}} 
          \caption{\label{fig3}
          The figure shows the energy of the linear chain for the cases when: (no kicks) only the bath acts on the system, 
          (a) kicks done to qubit A, (b) kicks done to qubit B
          and (c) kicks done to qubit C. The dashed black lines show the average energy of the quasi stationary state.
          The parameters used are
          $\kappa=\pi/2$, $\tauk=\pi$, $\delta_B=0.5$, $\delta_C=0.25$, $\chi=0.15$, $\chi'=0.1$ and $\beta=0.1$.}
  \end{center}
\end{figure}

\begin{figure}[!htbp]
  \begin{center}
      \resizebox{85mm}{!}{\includegraphics{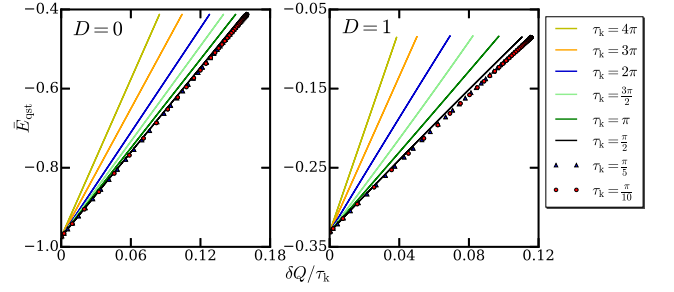}} 
          \caption{\label{fig4}
          Linear relation between the averaged energy of the system and the energy dissipated to the environment per period of kicks 
          at the quasi stationary regime when kicks are applied to qubit A. The figure shows the finite temperature and the zero
          temperature limit. The kick strength is varied from 0 to $2\pi$ and different periods of the kick have been used. 
          The parameters of the linear chain are: $\delta_B=0.5$, $\delta_C=0.25$, $\chi=0.15$, $\chi'=0.1$ and $\beta=0.1$.}
  \end{center}
\end{figure}

\begin{figure}[!htbp]
  \begin{center}
      \resizebox{85mm}{!}{\includegraphics{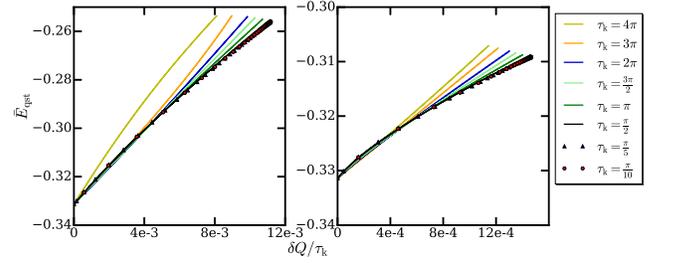}} 
          \caption{\label{fig5}
          Relation between the averaged energy of the system and the energy dissipated to the environment per period of kicks 
          at the quasi stationary regime when kicks are applied to qubit B (figure at the left) and C (figure at the right) for the finite temperature 
          limit with $D=1$. The kick strength is varied from 0 to $2\pi$ and different periods of the kick have been used. 
          The parameters of the linear chain are: $\delta_B=0.5$, $\delta_C=0.25$, $\chi=0.15$, $\chi'=0.1$ and $\beta=0.1$.}
  \end{center}
\end{figure}

The quasi stationary regime is reached by the system when it enters into 
a cycle limit dynamics where the amount of of energy dissipated 
to the environment between two consecutive kicks equals the amount
of energy received by the individual kicks. A profile of
the energy of the system at the time $\tau$; 
$E(\tau)=\langle \Hs \rangle$ is depicted in figure \ref{fig3}
for the zero temperature and finite temperature limits. In the figure 
one sees the quasi stationary is reached after certain time 
where the energy fluctuates around a constant value.
It has been shown for the kicked oscillator and the kicked rotor that at the 
quasi stationary regime, these systems follows a Fourier's law where the average energy 
of the systems and the dissipated energy to the environment per period of the kick are 
directly proportional, see~\eg~\cite{PrLoGo2016,LoGa2016}.
The averaged energy at the quasi stationary regime is defined as
\begin{equation}
\bar{E}_{\mt{qst}} = \lim_{n \rightarrow \infty}{E(\tau_n^{+})+E(\tau_n^{-})\over 2}
\end{equation}
where $\tau_n^{+}=\lim_{\delta \rightarrow 0 } n\tauk+\delta$ 
and $\tau_n^{-}=\lim_{\delta\rightarrow 0}n\tauk-\delta$
represents respectively the time immediately 
after and immediately before the n-th kick has happen. On the other hand, 
the dissipated energy per period of the kicks is defined as:
\begin{equation}\label{dq}
 \delta Q/\tauk =  \lim_{n \rightarrow \infty}{E(\tau_n^{+})-E(\tau_{n+1}^{-})\over \tauk}
\end{equation}
For our system, we have found a similar behavior for the linear chain as one can sees from 
figure~\ref{fig4} where the averaged energy at the quasi stationary regime
is plotted against the dissipated energy per period of the kick for kicks
applied to qubit A. 
As the period of the kicks becomes smaller (more frequent kicks) 
the proportionality of $\bar{E}_{\mt{qst}}$
to $\delta Q/\tauk$ becomes independent on the period of the kicks.
The independence on the of the slope to the
period of the kicks was observed in~\cite{PrLoGo2016}
for the kicked oscillator system where the slopes only depend on the damping rate. 
Nevertheless there is no fundamental reason why the slopes 
should not depend on the period of the kicks.
If the kicks applied to the other two qubits, 
a similar behavior is observed as in figure \ref{fig4}
for the zero temperature limit. Nevertheless 
we have found for the finite temperature limit, certain cases where
the relation does not seems to be linear anymore. This is shown 
in figure \ref{fig5} where we have depicted the finite temperature 
limit when the kicks are applied
to qubit B and qubit C.\\

\begin{figure}[!htbp]
  \begin{center}
      \resizebox{85mm}{!}{\includegraphics{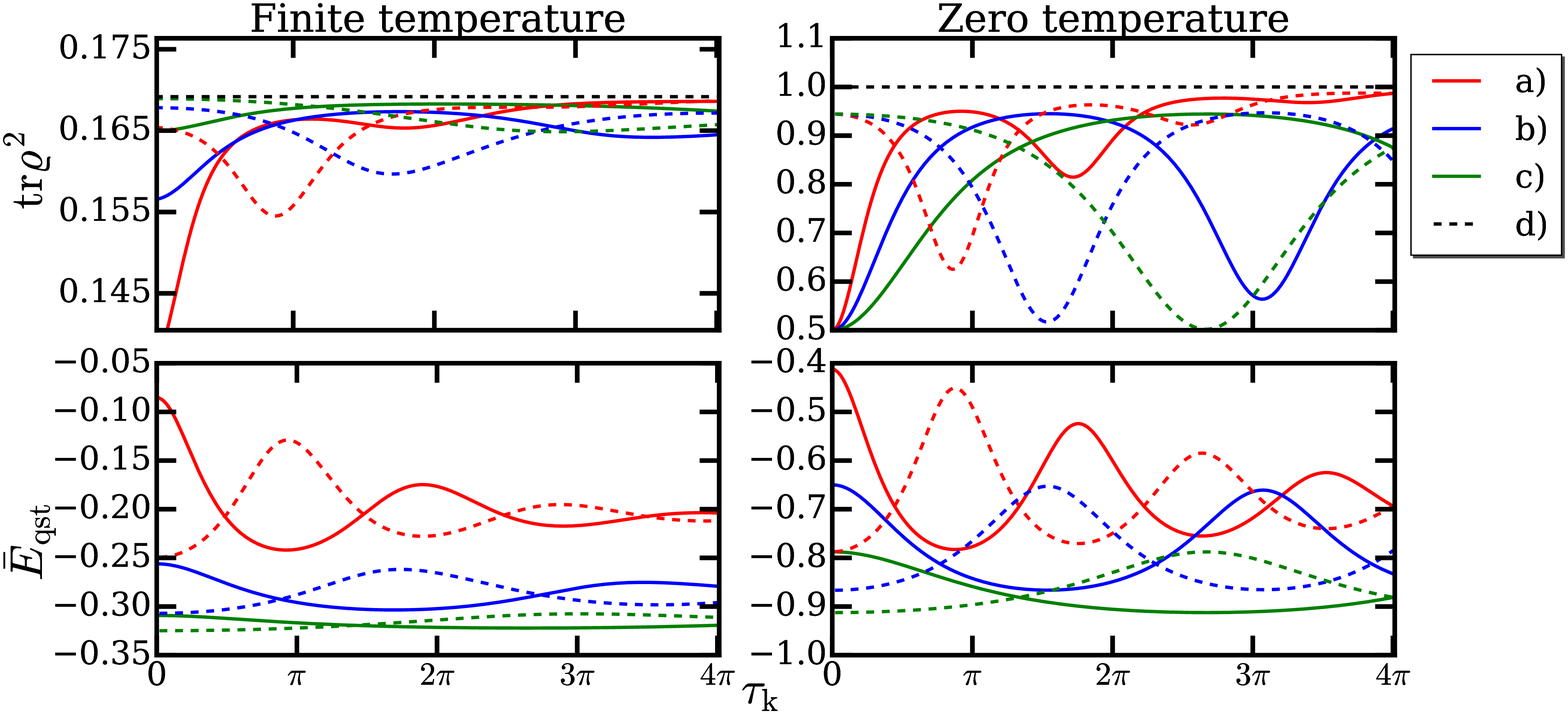}} 
          \caption{\label{fig6}
          Purity of the system at the quasi stationary regime for the finite temperature
          ($D=1$) and the zero temperature ($D=0$) limits as a function 
          of the period of the kicks when the kicks have been applied to a) qubit A, b) qubit B and c) qubit C.
          The continuous lines shows the purity when the kicks are done by an angle of $\pi/2$ around the $x$ axis 
          and the dashes lines shows the purity when the kicks are done in a $\pi$ angle around the $x$ axis and $\pi/2$ 
          angle around the $y$ axis. The dashed black lines labeled by d) represent the purity of the stationary state when only the bath acts 
          on the system. The parameters of the linear chain are: $\delta_B=0.5$, $\delta_C=0.25$, $\chi=0.15$, $\chi'=0.1$ and $\beta=0.1$.}
  \end{center}
\end{figure}

Now we place our attention back to the 
coherences appearing at the quasi stationary states. 
The coherent terms are a consequence of the angle of rotation, $\kappa=\pi/2$, 
for which the kick instantaneously changes the states of the kicked qubits into a superposition 
of states, \eg~if $n$th qubit is initially found in the state $|\psi\rangle_n= |0\rangle_n$,
then the application of a kick into the $x$-direction will yield: 
$R_{x,n}^{\pi /2} | 0 \rangle_n = 1/\sqrt{2}( |0\rangle_n + \rmi |1\rangle_n )$.
This superposition is kept in the system at a certain degree when the
quasi steady condition is reached because of the repeated application
of the kicks.
There are other ways to generate superposition of states as a quasi stationary 
condition. One possible way is to apply first a kick 
corresponding to a rotation of $\pi$ along the $x$ axis and afterwards to apply 
a kick corresponding to a rotation of $\pi/2$ along the $y$ axis. 
This will change the state of the qubit $|\psi\rangle_n= |0\rangle_n$ 
into $R_{x,n}^{\pi}R_{y,n}^{\pi /2} | 0 \rangle_n = \rmi/\sqrt{2}( |0\rangle_n  +  |1\rangle_n )$
which is the same superposed state but with a global constant phase.\\
Although a certain amount of coherence is gained in both cases, in general, 
the purity of the linear chain does not gets improved because the rest of the qubits
of the linear chain are subjet to the influence of the bath alone producing decoherence 
on them. In figure \ref{fig6} the purity, (first row), is plotted against the 
period of the kicks, for the cases where the kicks are done in the $x$ direction with 
an angle of $\pi/2$ (continuous lines) and when the kicks are done by an angle of $\pi$ in the $x$ direction 
and $\pi/2$ in the $y$ direction (dashed lines). 
One sees in the figure that the purity has a strong dependence on the period and the direction of the kicks
presenting some local maximums and minimums at different periods of the kicks. 
At the second row of figure \ref{fig6}, the average energy at the quasi steady regime 
is plotted and it also increases and decreases as function of the period of the kicks
meaning that the system passes through resonant and
non-resonant regions. The resonant regions coincide with the periods 
of smaller purity and vice versa. 

\section{\label{simkicks}Simultaneous kicking and the emergence of stationary entanglement}
\begin{figure}[!htbp]
  \begin{center}
      \resizebox{85mm}{!}{\includegraphics{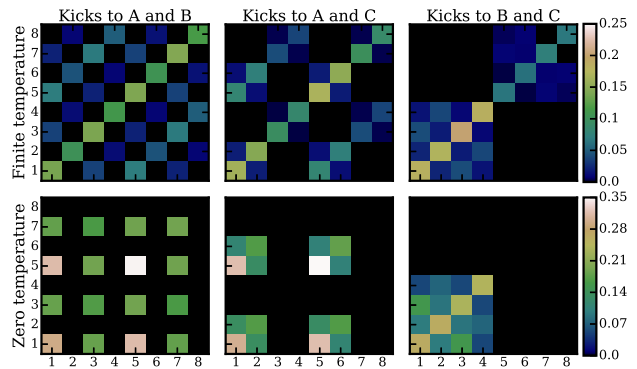}} 
          \caption{\label{fig7}
          Density matrix at the quasi stationary regime for the finite temperature ($D=1$) and the zero temperature ($D=0$) 
          for simultaneous kicking. The period of the kick used was {\color{red}$\tauk=4\pi$} and the kick strength was settled to $\pi/2$. 
          The parameters of the linear chain are: $\delta_B=0.5$, $\delta_C=0.25$, $\chi=0.15$, $\chi'=0.1$ and $\beta=0.1$.}
  \end{center}
\end{figure}

\begin{figure}[!htbp]
  \begin{center}
      \resizebox{85mm}{!}{\includegraphics{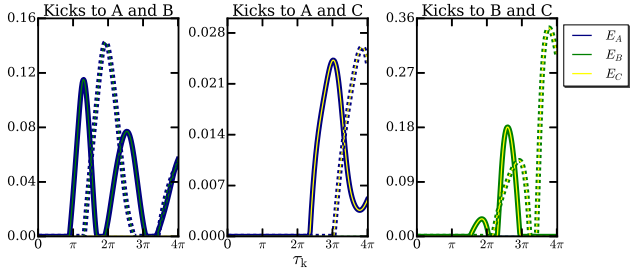}} 
          \caption{\label{fig8}
          Logarithmic negativity of the qubits at the zero temperature limit ($D=0$) 
          for simultaneous kicking as a function of the period of the kicks 
          for the cases when the kicks are applied in the $x$ direction with an angle $\pi/2$ to two 
          different qubits (continuous lines) and when the kicks are applied
          to one qubit first in the $x$ direction by an angle $\pi$ and afterwards in the $y$ direction 
          by an angle $\pi/2$ while the other qubit is kicked in the $x$ direction by an angle $\pi/2$. 
          The parameters of the linear chain are: $\delta_B=0.5$, $\delta_C=0.25$, $\chi=0.15$, $\chi'=0.1$ and $\beta=0.1$.}
  \end{center}
\end{figure}

\begin{figure}[!htbp]
  \begin{center}
      \resizebox{85mm}{!}{\includegraphics{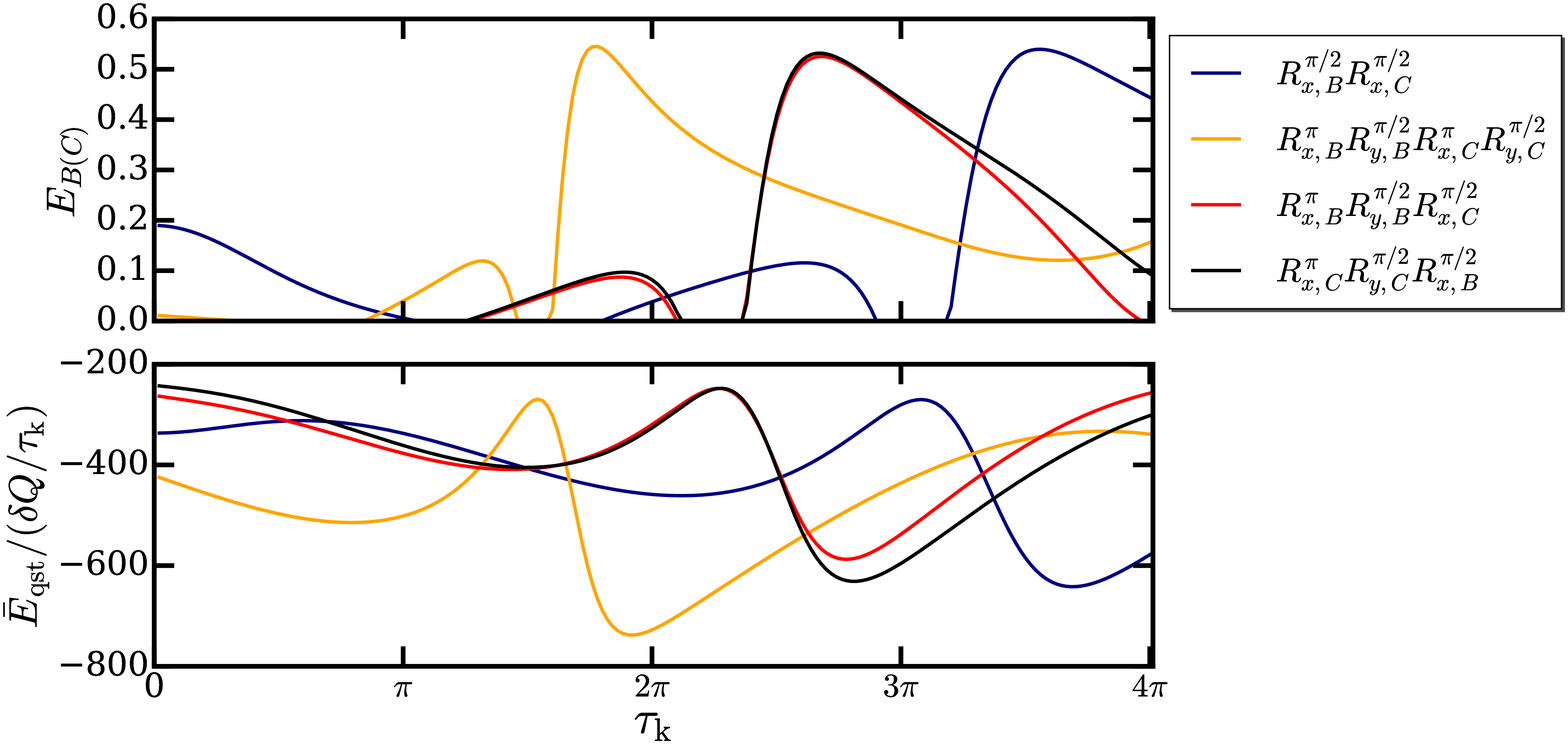}} 
          \caption{\label{fig9}
          At the first row are plotted the logarithmic negativities of the qubits B and C at the zero temperature limit ($D=0$) 
          for simultaneous kicking as a function of the period of the kicks with the kicks applied in different configurations 
          At the second row are plotted the ratio between the average energy and the dissipated energy as a function of the 
          period of the kicks for the correspondent cases considered at the first row. 
          The parameters of the linear chain are: $\delta_B=0.26$, $\delta_C=0.25$, $\chi_{AB}=0.011$, $\chi_{AC}=0.1$,
          $\chi_{BC}=0.15$ and $\beta=0.1$.}
  \end{center}
\end{figure}

Now we consider the scenario when simultaneous kicks
are applied to different qubits. In this case, the application of the kicks 
produces non-local changes on the system which together with 
the effects of the bath into the system it is possible to obtain 
a certain degree of entanglement between the kicked qubits as 
a stationary condition.
The entanglement produced among the qubits would be inherently resilient to
the effects of the environment in the sense that the
environment together with the application of kicks are the mechanism
that produce it. 
We begin by showing in figure \ref{fig7} the density matrices 
at the quasi stationary regime, when the kicks are simultaneously applied 
to two different qubits. In the figure, the kicks are applied 
in the $x$ direction with an angle of $\kappa=\pi/2$, such 
that the unitary operator representing the kicks is:
$R_{x,i}^{\pi/2}R_{x,j}^{\pi/2}$ with $i \neq j$ labeling the 
different qubits.
The pattern formed at the quasi stationary state contains new superpositions of states
which are the result of both mechanisms of kicks and dissipation,
\eg,~in the sub figure at the bottom left, which shows the case for kicks done to 
qubits A and B at the zero temperature limit, the bath is always driving the system to the
ground state which becomes the most likely state to be populated. On the other hand, 
the application of the kicks 
will have more influence over this state than any other, bringing the system 
into a non pure superposition states similar to: 
$|\psi\rangle\sim c_1 |000\rangle + c_3 |010\rangle  + c_5|100\rangle + c_7|110\rangle$. 
This superposition pattern does not corresponds to a pure state because the environment
is always acting on the system producing decoherence. Moreover, the decoherence makes the 
states of the system to be non separable and thus, a certain amount of entanglement 
between the qubits is induced. In order to measure the degree of entanglement we use
the logarithmic negativity~\cite{Plenio2005} defined as:
\begin{equation}
 E_j(\varrho)=\log_2\left(2\mathcal{N}_j+1\right)
\end{equation}
where $\mathcal{N}_j$ is the negativity of the $j$-th qubit defined as:
\begin{equation}
 \mathcal{N}_j = \sum_i{|\lambda_{ij}|-\lambda_{ij}\over 2}
\end{equation}
and $\lambda_{ij}$ are eigenvalues of the partial transpose $\varrho^{\Gamma_j}$ of $\varrho$, with 
respect to the $j$ qubit. The logarithmic negativity will measure how much entangled is the $j$th qubit 
with the rest of the system. In figure \ref{fig8} we show the logarithmic negativity
for the cases shown in figure \ref{fig7} (continuous lines), and another configuration of the kicks 
represented by the unitary operator $R_{x,j}^{\pi}R_{y,j}^{\pi/2}R_{x,i}^{\pi/2}$ with $i \neq j$ labeling the
different kicked qubits, this is; one qubit is first kicked in the $x$ direction by an angle $\pi$ and afterwards in the $y$ direction 
by an angle $\pi/2$ while the other qubit is kicked in the $x$ direction 
by an angle $\pi/2$. This configuration is chosen because it produces
higher rates of entanglement for certain periods of the kicks although, there might
exist some other configurations of the kicks producing larger rates of entanglement between the qubits.
Additionally, we only present the zero temperature limit case 
since for the finite temperature limit we have not found any entanglement
between the qubits in the parameter regime explored so far.  
In figure \ref{fig8}, one can observe that the largest rate of entanglement happens for the kicks done 
to qubit B and C which have a closer Larmor frequency among them. This suggest us one possible way to enhance entanglement 
by changing the configuration of the system, particularly 
by doing the Larmor frequencies of the kicked qubits, closer to each other.
This can be physically realizable by letting the qubits to lie closer in the linear chain,
since their Larmor frequencies are position dependent due to the gradient of the magnetic field field. 
In figure \ref{fig9} we have plotted at the first row the logarithmic entanglement of qubits B and C
as function of the period of the kicks, when kicks are done using different configurations and with the 
system parameters settled to:
$\delta_B=0.26$, $\delta_C=0.25$, $\chi_{AB}=0.011$, $\chi_{AC}=0.1$, $\chi_{BC}=0.15$. Now 
we have independently defined the Ising interaction rate according to the distance between the qubits. 
Additionally at the second row we have plotted the ratio of the average energy at the quasi stationary state $\bar{E}_{\mt{qst}}$ 
and the dissipated energy to the environment per period of the kick $\delta Q/\tauk$ versus the period of the kick.  
Figure \ref{fig9} shows that the maximum entanglement happens for periods of the kicks which 
corresponds to the global minimum of the ratio between the average energy and the dissipated 
energy per period of the kick. On the other hand, under the Fourier's law assumption~\cite{PrLoGo2016},
the ratio between the average energy and the dissipated energy is equivalent in definition to a
period dependent Fourier's coefficient $\mathcal{F}(\tauk)$ which describes the
proportionality rate between the difference of temperatures between the system and the bath, and the energy
exchange (dissipated energy) between the two systems $\Delta T = \mathcal{F}(\tauk) \delta Q /\tauk$ (notice 
that at the zero temperature limit this difference corresponds only to the average energy of the system, $\Delta T\sim \bar{E}_{\rm{qst}}$).
From the figure one notices that the Fourier's coefficient and the entanglement seems to follow an inverse relation 
such that the maximum rates of entanglement between the qubits appear for periods of the kicks where the Fourier's coefficient
is minimum. 
We should mention that the formation of entangled states 
between certain qubits in a linear chain by means of periodic kicks 
has been explored before in~\cite{Sainz2011}, nevertheless in this case,
the entanglement does not appears as a stationary condition and the interaction with the environment 
in fact will destroy it. In our case the formation of the stationary entanglement
appears in the system due to the collective action of the mechanisms of dissipation and non local kicks in the linear chain.
It may be possible to understand the formation of the entanglement as an 
emergent property of the system as it can be measured and classified as a property of a set 
of parts of the system which are in this case the two kicked qubits.

\section{\label{sum} Summary}
We have described the quasi stationary condition reached by 
a linear chain made of qubits 
subject to periodic kicks and dissipation.
The linear chain we have used 
has been a theoretical model for a certain 
type of quantum computing models.
For doing our study we have derived a master equation 
for which the degree of interaction to the environment 
depends on the energy of the different states of the linear chain. 
This model of dissipation leads to a stationary condition which corresponds 
to a Gibbs distribution at the finite temperature limit and 
to the ground state at the zero temperature limit which are the limits
one would expect.  We have described the conditions and the 
attributes of the non-equilibrium stationary states reached by the system  
when periodic delta kicks are applied to the 
qubits in two different situations: kicks applied to single qubits and
simultaneous kicks applied to the qubits. In the case of single kicked qubits,
we have found an endurable condition of the system to remain 
in a superposition state regardless of the effects of
the bath since the bath itself plays a crucial role in the formation of these 
states. Nevertheless we have found that the overall purity of the system does not 
gets improved since the rest of the linear chain remains under the influence 
of the bath. Also we have found resonant periods of the kicks for which
the degree of super position and the average energy of the system increases.
In the second case we have found the emergence of stationary entanglement when 
simultaneous kicks are applied to a pair of qubits of the linear chain. 
We have enhanced the rates of entanglement by changing the configuration of the 
system making the two kicked qubits to lie closer to each other and we observed 
that there exist an inverse relation between the entanglement and the Fourier's 
coefficient of the system at the quasi stationary regime.

\acknowledgments
We thank Thomas Gorin for the enlightening and useful discussions. We acknowledge
the hospitality of the Centro Internacional de Ciencias, UNAM where some of the
discussions took place.

\appendix
\section{\label{app} Derivation of a master equation}
In~\cite{LoLo2012a} a derivation of a master equation for
a linear chain of three nuclear spins system with second neighbor Ising 
interaction has been done and also similar lines of derivation 
of a master equation has been done for the quantum planar rotor
in~\cite{LoGa2016}. In both cases, the energy spectrum of the system 
has a non-equidistant spectrum and are coupled to the environment through 
creation and annihilation operators producing spontaneous emission and 
thermally induced processes. Here we follow similar
lines of the derivation of both cases to derive the master equation that
will account for our model. This master equation is not in a
Lindblad form but rather it is derived by 
Redfield approximations which we consider to work better for the
description of the spontaneous emission and thermally induced process
on a system with an non-equidistant spectrum. We start by writing 
the full Hamiltonian of the composite in the form
$H(t) = H_{\mt{c}} + W_{\mt{int}}(t)$ with 
$H_{\mt c} = \Hs + \Henv$ where $\Henv=\sum_{i}^{\infty}\hbar\omega_i a^{\dag}_ia_i,$
and $W_{\mt{int}}(t)= \Hrf(t) + \Hint$. 
The dynamical equation of the reduced density matrix for the spin chain system 
with an initially decoupled state of the system-environment,
$\varrho=\varrho_{\mt{s}}\otimes\sigma_{\mt{env}}$,
in the interaction picture with respect to $H_{\mt{c}}$, and
under the Born-Markov limit~\cite{BrePet02} can be
written in the following form:
\begin{eqnarray}\label{dme1}
{\rmd \tilde{\varrho}_{\mt s}(t)\over \rmd t } &=& {1\over \rmi \hbar}[\tilde{H}_{\mt{rf}}(t),\tilde{\varrho}_{\mt s}(t)]\\\nonumber
&& \!\!-{1\over \hbar^2 }\!\! \int_{0}^{\infty}\!\!\!\!\! \rmd \tau
\mt{Tr}_{\mt{e}} [\tilde{H}_{\mt {int}}(t),[\tilde{H}_{\mt {int}}(t-\tau),\tilde{\varrho}_{\mt s}(t)
\otimes\sigma_{\mt e}]].
\end{eqnarray}
The operators $\tilde{s}_j^{\pm}$ for $j=1,...,N$, in the 
interaction picture have the form:
\begin{equation}\label{deink}
\tilde{s}_j^{\pm}(t)=s^{\pm}_je^{\pm i\h{\Omega}_j t},
\end{equation}
where 
\begin{equation}\label{omop}
\h{\Omega}_j=\omega_j+{J\over \hbar}(s_{j+1}^z+s_{j-1}^z)+{J'\over \hbar}(s_{j+2}^z+s_{j-2}^z), 
 \end{equation}
is an frequency operator that commutes with the Hamiltonian $\Hs$ and whose eigenvalues are
the transition frequencies of the different states.
The interaction Hamiltonian between the spin chain and the environment
is represented by a coupling between the polarization operator and a 
Bosonic modes operators. 
Since the baths are supposed to be in
a stationary Boltzmann states: $\sigma_{\mt {e}} = 
\prod_{k}{1\over \mathcal{Z}} \sum_{n_k} \rme^{-E_{n_k}/k_BT}|n_k\ra\la n_k | $,
any perturbation thermalizes immediately and also the
and also the self correlation functions of the baths are null: 
$\la \h{a}_i^{\dag}(s)\h{a}_k^{\dag}(t)\ra=\la \h{a}_i(s)\h{a}_k(t)\ra = 0$.
The bath correlation functions appearing in (\ref{dme1}) have the 
following form: 
\begin{eqnarray}\label{corffa1}
 \sum_{i,k} g_{ij}g\p*_{kl}\mathcal{C}_{ik}(\tau) &=&
\sum_i|g|^2_{ijl}\rme^{-\rmi \omega_i \tau}\left(N(\omega_i)+1\right)\\\label{corffa2}
\sum_{i,k} g\p*_{ij}g_{kl}\mathcal{C}\p*_{ik}(\tau) &=&
\sum_i|g|^2_{ijl}\rme^{i\omega_i\tau}N(\omega_i),
\end{eqnarray}
where $\mathcal{C}_{ik}(\tau)=\langle a_i(\tau)a^{\dag}_k\rangle$, 
$\mathcal{C}_{ik}(\tau)\p* =\langle a_i^{\dag}(\tau)a_k\rangle$ and 
$N(\omega_i)=\left(\rme^{\omega_i\hbar/k_BT}-1\right)\p{-1}$
are the Planck's distribution function.
We assume the sum over $i$ is dense (there are an uncountable number of 
radiation modes) and the continuous limit can be taken. The number of 
characteristic frequencies with wave vector components $\vec{f}$ in 
the interval $\rmd f_x \rmd f_y \rmd f_z$ in the volume $V$ is given by 
$V 4\pi f^2\rmd f /(2\pi)^3 = {V\omega^2}\rmd\omega/\pi^2 c^3$,
 where $f=c\cdot \omega$.  Thus the sum in the correlation functions 
 can be changed by an integration over the frequencies
 with the proper weight factor,
\begin{eqnarray}\label{corrfb1}
 \sum_{i,k} g_{ij}g\p*_{kl}\mathcal{C}_{ik}(\tau) &=& {\gamma_{jl}\over \pi}\int_{-\infty}^{\infty}
\!\!\!\!\rmd\omega \,\omega^3 (N(\omega)+1)e^{ - \rmi\omega\tau}\\\label{corrfb2}
\sum_{i,k} g\p*_{ij}g_{kl}\mathcal{C}\p*_{ik}(\tau) &=& 
{\gamma_{jl}\over \pi}\int_{-\infty}^{\infty}\!\!\!\! \rmd\omega \,\omega^3 N(\omega)
e^{\rmi\omega\tau}
\end{eqnarray}
where $\gamma_{jl}={V|g|^2_{jl}/\pi c^3}$.
The correlation functions becomes the Fourier transform of a spectral density
associated to the continuous  modes in the thermal bath
and we have assumed a linear dependence on the characteristic frequencies 
of the radiation modes, $|g|_{ijl}^2=|g|^2_{jl}\omega_i$.
By writing equation (\ref{dme1}) back in the Schr\"odingers picture
and using (\ref{corrfb1}) and (\ref{corrfb2}), we write for the master equation: 
\begin{eqnarray}\label{dme2}
{\rmd  \varrho_{\mt s} (t)\over \rmd t } &=& {1\over \rmi \hbar}[\Hs+\Hrf(t), \varrho_{\mt s}]\\\nonumber
&& \!\!-{1\over \hbar^2 }\!\!\sum_{j,l=1}\p{N}{\gamma_{jl}\over \pi}
\int_{0}^{\infty}\!\!\!\!\! \rmd \tau\!\!
\int_{-\infty}^{\infty}\!\!\!\!\! \rmd \omega\, \omega\p{3}\,\mathcal{R}_{j,l}(\omega,\tau)[\varrho_{\mt{s}}]
\end{eqnarray}
where the super operator $\mathcal{R}_{j,l}(\omega,\tau)[\varrho_{\mt{s}}]$ is defined as
\begin{eqnarray}\nonumber
 \mathcal{R}_{j,l}(\omega,\tau)[\varrho_{\mt{s}}]&=& \left(N(\omega)+1\right)\rme\p{-\rmi\left(\omega-\h{\Omega}_l\right)\tau}
 \mathrm{D}_1[\varrho_{\mt s}]\\
 && + N(\omega)\rme\p{\rmi\left(\omega-\h{\Omega}_l\right)\tau}
 \mathrm{D}_2[\varrho_{\mt s}]+h.c.
\end{eqnarray}
with $\mathrm{D}_1[\varrho_{\mt s}]= s_j\p{+}s_l\p{-}\varrho_{\mt s} - s_l\p{-}\varrho_{\mt s}s_j\p{+}$
and $\mathrm{D}_2[\varrho_{\mt s}]= s_j\p{-}s_l\p{+}\varrho_{\mt s} - s_l\p{+}\varrho_{\mt s}s_j\p{-}$. 
Now we can exchange the order of integration in (\ref{dme2}) and evaluate the integrals by introducing a full 
eigenbasis of $\Hs$, lets say $\One = \sum_{m}|m\rangle\langle m|$ and call $\Omega_{lm}$, the eigenvalues of 
the operator $\h{\Omega}_l$, ($\h{\Omega}_l|m\rangle = \Omega_{lm}|m\rangle$), for the jth spin.
For the $\tau$ integration we can 
separate the real and the imaginary part by using the known relation
$\int_{0}^{\infty} \rmd\tau \rme^{\pm \rmi \epsilon \tau} = \pi \delta(\epsilon)\mp \rmi {\mathbf{P} / \epsilon}$, 
where $\mathbf{P}$ is the Cauchy's principal value. For the real part, integration over $\tau$
will yield delta functions of the form $\delta(\omega-\Omega_{lm})$.
Consequently, integration over $\omega$ will yield :
$\int_{ -\infty }^{ \infty } \, \rmd \omega\,   \delta \left(\omega - \Omega_{lm}\right)
\omega\p{3} N\left(\omega\right)
 =\Omega\p{3}_l\, N(\Omega_{lm})$. This real part is responsible of the non-unitary dynamics of the system 
 yielding the dissipative processes and thermalization processes. On the other hand, the imaginary part 
 contain some non physical contributions to the dynamics that can be solved if we
 neglect a small term under the assumption of $\omega_l \gg J (J')\hbar$ and additionally assume the 
 secular approximation which is equivalent to consider $\gamma_{ij}=\gamma_i\delta_{ij}$. With this assumptions 
 the imaginary term can be incorporated to the von Neumann dynamics.
 By recovering the identity we write for the master equation:
\begin{equation}\label{dme3}
\rmi \hbar {\rmd \varrho_{\mt s}(t)\over \rmd t } = [\Hs+\Hrf(t)+H_{LS}, \varrho_{\mt s}]
+{\rmi\over \hbar} \mathcal{D}[\varrho_{\mt s}]\\\nonumber
\end{equation}
 where 
 \begin{eqnarray}\nonumber
\mathcal{D}[\varrho_{\mt s}] &=&
- \sum_{l=1}\p{N} \left(\left[ \h{\mt{O}}\p{(1)}_{l}\!(T)\,s_l\p{+},s_l\p{-}\varrho_{\mt s}\right]
+\left[ \varrho_{\mt s}s_l\p{+},s_l\p{-}\,\h{\mt{O}}\p{(1)}_{l}\!(T)\right]\right.\\\nonumber
&&+\left.\left[ \h{\mt{O}}\p{(2)}_{l}\!(T)\,s_l\p{-},s_l\p{+}\varrho_{\mt s}\right]
+\left[ \varrho_{\mt s}s_l\p{-},s_l\p{+}\,\h{\mt{O}}\p{(2)}_{l}\!(T)\right]\right)\\\label{dis}
\end{eqnarray}
with 
\begin{eqnarray}\label{GT1}
\h{\mt{O}}\p{(1)}_{l} (T) &=& \gamma_l \h{\Omega}_l\p{3}\left(N(\h{\Omega}_l,T)\!+\!1\right)\,,\\\label{GT2}
 \h{\mt{O}}\p{(2)}_{l} (T) &= &\gamma_l \h{\Omega}_l\p{3} N(\h{\Omega}_l,T),
 \end{eqnarray}
 and
 \begin{equation}\label{NT}
N(\h{\Omega}_l,T)=\left(\rme^{
{\h{\Omega}_l\hbar/ k_B T}}-1\right)\p{-1}  
 \end{equation}
with $\h{\Omega}_l$ given by (\ref{omop}). The new term included in 
the von Neumann dynamics, $H_{LS}$ is:
 \begin{equation}\label{hls}
H_{LS}= \sum_{l=1}\p{N}\left( \h{\Gamma}\p{(1)}_{l}(T) s_{l}\p{+}s_{l}\p{-}+ \h{\Gamma}\p{(2)}_{l}(T) s_{l}\p{-}s_{l}\p{+}  \right)
 \end{equation}
with
 \begin{eqnarray}\label{GT11}
  \h{\Gamma}\p{(1)}_{l}(T) &=& {\gamma_{l}\over \pi\hbar }\int_{-\infty}\p{\infty} 
  \rmd \omega {\omega\p{3}\left(N(\omega)+1\right)\over \omega - \h{\Omega}_l},\\\label{GT22}
  \h{\Gamma}\p{(2)}_{l}(T) &=& {\gamma_{l}\over \pi \hbar}\int_{-\infty}\p{\infty} 
  \rmd \omega {\omega\p{3}N(\omega)\over \omega - \h{\Omega}_l}.  
 \end{eqnarray}
 The term $\mathcal{D}[\varrho_{\mt s}]$ in (\ref{dme3})
describes spontaneous emission and thermally induced process which occur at
a rate that depends on the energy level distribution of the spin chain and the correlation 
of these process for the different spins.
 The transition probabilities of the system due to the
spontaneous emission process occur with rates that depends
on the cubic power of the energy level difference of each spin, 
$\approx \gamma_l \h{\Omega}\p{3}_{l}$ while the
probability of increasing energy states due to
the thermally induced processes occur with a rate of 
$\gamma_{l}\h{\Omega}\p{3}_{l}N(\h{\Omega}_{l})$
which decays exponentially for large energy states.
 On the other hand the term $H_{LS}$ in (\ref{hls}) commutes with the Hamiltonian of the system 
and contributes with a certain shift to the eigen energies of the system. 
Typically this term is related to a Lamb shift effect and sometimes is simply neglected.
This will be our case since we want to focus only on the non unitary dynamics effects of the bath. 
The  zero temperature limit 
is considered when the temperature of the bath is sufficiently small compared
to the energy transitions of the linear chain and one can do the limit $T\rightarrow 0$ in the
operators (\ref{GT1}) and (\ref{GT2}) with (\ref{NT}). In this case,  
the dissipative term of the master equation describes a pure spontaneous emission process
and the super operator responsible of the dissipation $\mathcal{D} [\varrho_{\mt s}]$ takes the form:
\begin{eqnarray}\nonumber
\mathcal{D} [\varrho_{\mt s}] &=&
-\sum_{l=1}\p{N} \gamma_l \left\{\left[\h{\Omega}\p{3}_{l}\,s_l\p{+},s_l\p{-}\varrho_{\mt s} \right]
+\left[ \varrho_{\mt s} \,s_l\p{+},s_l\p{-}\,\h{\Omega}\p{3}_{l}\right]\right\}.
\end{eqnarray}
At the finite temperature limit, the system reaches a stationary state which is a Gibbs distribution 
mixture of states and
as the temperature increases the states get closer together until it reaches an homogeneous mixture for infinite temperatures.
At the zero temperature limit, the system reaches a stationary state which is a pure state as in the 
spontaneous emission process where all the states become populated during the transients and in the long time limit
only the ground state becomes populated.

\bibliographystyle{apsrev4-1}
\bibliography{man}

\begin{thebibliography}{30}%
\makeatletter
\providecommand \@ifxundefined [1]{%
 \@ifx{#1\undefined}
}%
\providecommand \@ifnum [1]{%
 \ifnum #1\expandafter \@firstoftwo
 \else \expandafter \@secondoftwo
 \fi
}%
\providecommand \@ifx [1]{%
 \ifx #1\expandafter \@firstoftwo
 \else \expandafter \@secondoftwo
 \fi
}%
\providecommand \natexlab [1]{#1}%
\providecommand \enquote  [1]{``#1''}%
\providecommand \bibnamefont  [1]{#1}%
\providecommand \bibfnamefont [1]{#1}%
\providecommand \citenamefont [1]{#1}%
\providecommand \href@noop [0]{\@secondoftwo}%
\providecommand \href [0]{\begingroup \@sanitize@url \@href}%
\providecommand \@href[1]{\@@startlink{#1}\@@href}%
\providecommand \@@href[1]{\endgroup#1\@@endlink}%
\providecommand \@sanitize@url [0]{\catcode `\\12\catcode `\$12\catcode
  `\&12\catcode `\#12\catcode `\^12\catcode `\_12\catcode `\%12\relax}%
\providecommand \@@startlink[1]{}%
\providecommand \@@endlink[0]{}%
\providecommand \url  [0]{\begingroup\@sanitize@url \@url }%
\providecommand \@url [1]{\endgroup\@href {#1}{\urlprefix }}%
\providecommand \urlprefix  [0]{URL }%
\providecommand \Eprint [0]{\href }%
\providecommand \doibase [0]{http://dx.doi.org/}%
\providecommand \selectlanguage [0]{\@gobble}%
\providecommand \bibinfo  [0]{\@secondoftwo}%
\providecommand \bibfield  [0]{\@secondoftwo}%
\providecommand \translation [1]{[#1]}%
\providecommand \BibitemOpen [0]{}%
\providecommand \bibitemStop [0]{}%
\providecommand \bibitemNoStop [0]{.\EOS\space}%
\providecommand \EOS [0]{\spacefactor3000\relax}%
\providecommand \BibitemShut  [1]{\csname bibitem#1\endcsname}%
\let\auto@bib@innerbib\@empty
%</preamble>
\bibitem [{\citenamefont {Haken}(1970)}]{Haken1970}%
  \BibitemOpen
  \bibfield  {author} {\bibinfo {author} {\bibfnamefont {H.}~\bibnamefont
  {Haken}},\ }\enquote {\bibinfo {title} {Laser theory},}\ in\ \href {\doibase
  10.1007/978-3-662-22091-7_1} {\emph {\bibinfo {booktitle} {Light and Matter
  Ic / Licht und Materie Ic}}},\ \bibinfo {editor} {edited by\ \bibinfo
  {editor} {\bibfnamefont {L.}~\bibnamefont {Genzel}}}\ (\bibinfo  {publisher}
  {Springer Berlin Heidelberg},\ \bibinfo {address} {Berlin, Heidelberg},\
  \bibinfo {year} {1970})\ pp.\ \bibinfo {pages} {1--304}\BibitemShut {NoStop}%
\bibitem [{\citenamefont {Haken}(2004)}]{Haken2004}%
  \BibitemOpen
  \bibfield  {author} {\bibinfo {author} {\bibfnamefont {H.}~\bibnamefont
  {Haken}},\ }\href@noop {} {\emph {\bibinfo {title} {Synergetics, Introduction
  and Advanced Topics}}}\ (\bibinfo  {publisher} {Springer},\ \bibinfo {year}
  {2004})\BibitemShut {NoStop}%
\bibitem [{\citenamefont {Lutz}\ and\ \citenamefont
  {Weidenmüller}(1999)}]{LutWei99}%
  \BibitemOpen
  \bibfield  {author} {\bibinfo {author} {\bibfnamefont {E.}~\bibnamefont
  {Lutz}}\ and\ \bibinfo {author} {\bibfnamefont {H.~A.}\ \bibnamefont
  {Weidenmüller}},\ }\href@noop {} {\bibfield  {journal} {\bibinfo  {journal}
  {Physica A}\ }\textbf {\bibinfo {volume} {267}},\ \bibinfo {pages} {354}
  (\bibinfo {year} {1999})}\BibitemShut {NoStop}%
\bibitem [{\citenamefont {Gorin}\ and\ \citenamefont {Seligman}(2002)}]{GS02b}%
  \BibitemOpen
  \bibfield  {author} {\bibinfo {author} {\bibfnamefont {T.}~\bibnamefont
  {Gorin}}\ and\ \bibinfo {author} {\bibfnamefont {T.~H.}\ \bibnamefont
  {Seligman}},\ }\href@noop {} {\bibfield  {journal} {\bibinfo  {journal} {J.
  Opt. B: Quantum Semiclass. Opt.}\ }\textbf {\bibinfo {volume} {4}},\ \bibinfo
  {pages} {S386} (\bibinfo {year} {2002})},\ \bibinfo {note} {topical issue:
  Mysteries and Paradoxes in Quantum Mechanics IV Quantum interference
  phenomena (Workshop held at Gargano, Italy, August 2001)}\BibitemShut
  {NoStop}%
\bibitem [{\citenamefont {Carrera}\ \emph {et~al.}(2014)\citenamefont
  {Carrera}, \citenamefont {Gorin},\ and\ \citenamefont
  {Seligman}}]{CaGoSe2014}%
  \BibitemOpen
  \bibfield  {author} {\bibinfo {author} {\bibfnamefont {M.}~\bibnamefont
  {Carrera}}, \bibinfo {author} {\bibfnamefont {T.}~\bibnamefont {Gorin}}, \
  and\ \bibinfo {author} {\bibfnamefont {T.~H.}\ \bibnamefont {Seligman}},\
  }\href {\doibase 10.1103/PhysRevA.90.022107} {\bibfield  {journal} {\bibinfo
  {journal} {Phys. Rev. A}\ }\textbf {\bibinfo {volume} {90}},\ \bibinfo
  {pages} {022107} (\bibinfo {year} {2014})}\BibitemShut {NoStop}%
\bibitem [{\citenamefont {Moreno}\ \emph {et~al.}(2015)\citenamefont {Moreno},
  \citenamefont {Gorin},\ and\ \citenamefont {Seligman}}]{MoGoSe2015}%
  \BibitemOpen
  \bibfield  {author} {\bibinfo {author} {\bibfnamefont {H.~J.}\ \bibnamefont
  {Moreno}}, \bibinfo {author} {\bibfnamefont {T.}~\bibnamefont {Gorin}}, \
  and\ \bibinfo {author} {\bibfnamefont {T.~H.}\ \bibnamefont {Seligman}},\
  }\href {\doibase 10.1103/PhysRevA.92.030104} {\bibfield  {journal} {\bibinfo
  {journal} {Phys. Rev. A}\ }\textbf {\bibinfo {volume} {92}},\ \bibinfo
  {pages} {030104} (\bibinfo {year} {2015})}\BibitemShut {NoStop}%
\bibitem [{\citenamefont {Langemeyer}\ and\ \citenamefont
  {Holthaus}(2014)}]{LanHol14}%
  \BibitemOpen
  \bibfield  {author} {\bibinfo {author} {\bibfnamefont {M.}~\bibnamefont
  {Langemeyer}}\ and\ \bibinfo {author} {\bibfnamefont {M.}~\bibnamefont
  {Holthaus}},\ }\href {\doibase 10.1103/PhysRevE.89.012101} {\bibfield
  {journal} {\bibinfo  {journal} {Phys. Rev. E}\ }\textbf {\bibinfo {volume}
  {89}},\ \bibinfo {pages} {012101} (\bibinfo {year} {2014})}\BibitemShut
  {NoStop}%
\bibitem [{\citenamefont {Ketzmerick}\ and\ \citenamefont
  {Wustmann}(2010)}]{KetWus10}%
  \BibitemOpen
  \bibfield  {author} {\bibinfo {author} {\bibfnamefont {R.}~\bibnamefont
  {Ketzmerick}}\ and\ \bibinfo {author} {\bibfnamefont {W.}~\bibnamefont
  {Wustmann}},\ }\href {\doibase 10.1103/PhysRevE.82.021114} {\bibfield
  {journal} {\bibinfo  {journal} {Phys. Rev. E}\ }\textbf {\bibinfo {volume}
  {82}},\ \bibinfo {pages} {021114} (\bibinfo {year} {2010})}\BibitemShut
  {NoStop}%
\bibitem [{\citenamefont {Prado~Reynoso}\ \emph {et~al.}(2017)\citenamefont
  {Prado~Reynoso}, \citenamefont {L\'opez~V\'azquez},\ and\ \citenamefont
  {Gorin}}]{PrLoGo2017}%
  \BibitemOpen
  \bibfield  {author} {\bibinfo {author} {\bibfnamefont {M.~A.}\ \bibnamefont
  {Prado~Reynoso}}, \bibinfo {author} {\bibfnamefont {P.~C.}\ \bibnamefont
  {L\'opez~V\'azquez}}, \ and\ \bibinfo {author} {\bibfnamefont
  {T.}~\bibnamefont {Gorin}},\ }\href {\doibase 10.1103/PhysRevA.95.022118}
  {\bibfield  {journal} {\bibinfo  {journal} {Phys. Rev. A}\ }\textbf {\bibinfo
  {volume} {95}},\ \bibinfo {pages} {022118} (\bibinfo {year}
  {2017})}\BibitemShut {NoStop}%
\bibitem [{\citenamefont {V\'azquez}\ and\ \citenamefont
  {Garc\'ia}(2016)}]{LoGa2016}%
  \BibitemOpen
  \bibfield  {author} {\bibinfo {author} {\bibfnamefont {P.~C.~L.}\
  \bibnamefont {V\'azquez}}\ and\ \bibinfo {author} {\bibfnamefont
  {A.}~\bibnamefont {Garc\'ia}},\ }\href
  {http://stacks.iop.org/1402-4896/91/i=5/a=055101} {\bibfield  {journal}
  {\bibinfo  {journal} {Physica Scripta}\ }\textbf {\bibinfo {volume} {91}},\
  \bibinfo {pages} {055101} (\bibinfo {year} {2016})}\BibitemShut {NoStop}%
\bibitem [{\citenamefont {Gardiner}\ \emph {et~al.}(1997)\citenamefont
  {Gardiner}, \citenamefont {Cirac},\ and\ \citenamefont {Zoller}}]{GCZ97}%
  \BibitemOpen
  \bibfield  {author} {\bibinfo {author} {\bibfnamefont {S.~A.}\ \bibnamefont
  {Gardiner}}, \bibinfo {author} {\bibfnamefont {J.~I.}\ \bibnamefont {Cirac}},
  \ and\ \bibinfo {author} {\bibfnamefont {P.}~\bibnamefont {Zoller}},\
  }\href@noop {} {\bibfield  {journal} {\bibinfo  {journal} {Phys. Rev. Lett.}\
  }\textbf {\bibinfo {volume} {79}},\ \bibinfo {pages} {4790} (\bibinfo {year}
  {1997})}\BibitemShut {NoStop}%
\bibitem [{\citenamefont {Billam}\ and\ \citenamefont
  {Gardiner}(2009)}]{BilGar09}%
  \BibitemOpen
  \bibfield  {author} {\bibinfo {author} {\bibfnamefont {T.~P.}\ \bibnamefont
  {Billam}}\ and\ \bibinfo {author} {\bibfnamefont {S.~A.}\ \bibnamefont
  {Gardiner}},\ }\href {\doibase 10.1103/PhysRevA.80.023414} {\bibfield
  {journal} {\bibinfo  {journal} {Phys. Rev. A}\ }\textbf {\bibinfo {volume}
  {80}},\ \bibinfo {pages} {023414} (\bibinfo {year} {2009})}\BibitemShut
  {NoStop}%
\bibitem [{\citenamefont {Dana}\ and\ \citenamefont
  {Dorofeev}(2005)}]{DaDo2005}%
  \BibitemOpen
  \bibfield  {author} {\bibinfo {author} {\bibfnamefont {I.}~\bibnamefont
  {Dana}}\ and\ \bibinfo {author} {\bibfnamefont {D.~L.}\ \bibnamefont
  {Dorofeev}},\ }\href {\doibase 10.1103/PhysRevE.72.046205} {\bibfield
  {journal} {\bibinfo  {journal} {Phys. Rev. E}\ }\textbf {\bibinfo {volume}
  {72}},\ \bibinfo {pages} {046205} (\bibinfo {year} {2005})}\BibitemShut
  {NoStop}%
\bibitem [{\citenamefont {Elyutin}\ and\ \citenamefont
  {Rubtsov}(2008)}]{El2008}%
  \BibitemOpen
  \bibfield  {author} {\bibinfo {author} {\bibfnamefont {P.~V.}\ \bibnamefont
  {Elyutin}}\ and\ \bibinfo {author} {\bibfnamefont {A.~N.}\ \bibnamefont
  {Rubtsov}},\ }\href {http://stacks.iop.org/1751-8121/41/i=5/a=055103}
  {\bibfield  {journal} {\bibinfo  {journal} {J. Phys. A}\ }\textbf {\bibinfo
  {volume} {41}},\ \bibinfo {pages} {055103} (\bibinfo {year}
  {2008})}\BibitemShut {NoStop}%
\bibitem [{\citenamefont {Fishman}\ \emph {et~al.}(1982)\citenamefont
  {Fishman}, \citenamefont {Grempel},\ and\ \citenamefont {Prange}}]{Fi1982}%
  \BibitemOpen
  \bibfield  {author} {\bibinfo {author} {\bibfnamefont {S.}~\bibnamefont
  {Fishman}}, \bibinfo {author} {\bibfnamefont {D.~R.}\ \bibnamefont
  {Grempel}}, \ and\ \bibinfo {author} {\bibfnamefont {R.~E.}\ \bibnamefont
  {Prange}},\ }\href {\doibase 10.1103/PhysRevLett.49.509} {\bibfield
  {journal} {\bibinfo  {journal} {Phys. Rev. Lett.}\ }\textbf {\bibinfo
  {volume} {49}},\ \bibinfo {pages} {509} (\bibinfo {year} {1982})}\BibitemShut
  {NoStop}%
\bibitem [{\citenamefont {Anderson}(1978)}]{An1978}%
  \BibitemOpen
  \bibfield  {author} {\bibinfo {author} {\bibfnamefont {P.~W.}\ \bibnamefont
  {Anderson}},\ }\href {\doibase 10.1103/RevModPhys.50.191} {\bibfield
  {journal} {\bibinfo  {journal} {Rev. Mod. Phys.}\ }\textbf {\bibinfo {volume}
  {50}},\ \bibinfo {pages} {191} (\bibinfo {year} {1978})}\BibitemShut
  {NoStop}%
\bibitem [{\citenamefont {Ad~Lagendijk}\ and\ \citenamefont
  {Wiersma}(2009)}]{Ad2009}%
  \BibitemOpen
  \bibfield  {author} {\bibinfo {author} {\bibfnamefont {B.~v.~T.}\
  \bibnamefont {Ad~Lagendijk}}\ and\ \bibinfo {author} {\bibfnamefont {D.~S.}\
  \bibnamefont {Wiersma}},\ }\href {http://dx.doi.org/10.1063/1.3206091}
  {\bibfield  {journal} {\bibinfo  {journal} {Physics Today}\ }\textbf
  {\bibinfo {volume} {62}} (\bibinfo {year} {2009})}\BibitemShut {NoStop}%
\bibitem [{\citenamefont {Berman}\ \emph {et~al.}(2000)\citenamefont {Berman},
  \citenamefont {Doolen}, \citenamefont {L\'opez},\ and\ \citenamefont
  {Tsifrinovich}}]{Be2000}%
  \BibitemOpen
  \bibfield  {author} {\bibinfo {author} {\bibfnamefont {G.~P.}\ \bibnamefont
  {Berman}}, \bibinfo {author} {\bibfnamefont {G.~D.}\ \bibnamefont {Doolen}},
  \bibinfo {author} {\bibfnamefont {G.~V.}\ \bibnamefont {L\'opez}}, \ and\
  \bibinfo {author} {\bibfnamefont {V.~I.}\ \bibnamefont {Tsifrinovich}},\
  }\href {\doibase 10.1103/PhysRevA.61.062305} {\bibfield  {journal} {\bibinfo
  {journal} {Phys. Rev. A}\ }\textbf {\bibinfo {volume} {61}},\ \bibinfo
  {pages} {062305} (\bibinfo {year} {2000})}\BibitemShut {NoStop}%
\bibitem [{\citenamefont {L\'opez}\ \emph {et~al.}(2003)\citenamefont
  {L\'opez}, \citenamefont {Quezada}, \citenamefont {Berman}, \citenamefont
  {Doolen},\ and\ \citenamefont {Tsifrinovich}}]{GV2003}%
  \BibitemOpen
  \bibfield  {author} {\bibinfo {author} {\bibfnamefont {G.~V.}\ \bibnamefont
  {L\'opez}}, \bibinfo {author} {\bibfnamefont {J.}~\bibnamefont {Quezada}},
  \bibinfo {author} {\bibfnamefont {G.~P.}\ \bibnamefont {Berman}}, \bibinfo
  {author} {\bibfnamefont {G.~D.}\ \bibnamefont {Doolen}}, \ and\ \bibinfo
  {author} {\bibfnamefont {V.~I.}\ \bibnamefont {Tsifrinovich}},\ }\href
  {http://stacks.iop.org/1464-4266/5/i=2/a=311} {\bibfield  {journal} {\bibinfo
   {journal} {J Opt B Quantum Semiclassical Opt}\ }\textbf {\bibinfo {volume}
  {5}},\ \bibinfo {pages} {184} (\bibinfo {year} {2003})}\BibitemShut {NoStop}%
\bibitem [{\citenamefont {Viola}\ and\ \citenamefont {Lloyd}(1998)}]{Vi1998}%
  \BibitemOpen
  \bibfield  {author} {\bibinfo {author} {\bibfnamefont {L.}~\bibnamefont
  {Viola}}\ and\ \bibinfo {author} {\bibfnamefont {S.}~\bibnamefont {Lloyd}},\
  }\href {\doibase 10.1103/PhysRevA.58.2733} {\bibfield  {journal} {\bibinfo
  {journal} {Phys. Rev. A}\ }\textbf {\bibinfo {volume} {58}},\ \bibinfo
  {pages} {2733} (\bibinfo {year} {1998})}\BibitemShut {NoStop}%
\bibitem [{\citenamefont {Rego}\ \emph {et~al.}(2009)\citenamefont {Rego},
  \citenamefont {Santos},\ and\ \citenamefont {Batista}}]{Re2009}%
  \BibitemOpen
  \bibfield  {author} {\bibinfo {author} {\bibfnamefont {L.~G.}\ \bibnamefont
  {Rego}}, \bibinfo {author} {\bibfnamefont {L.~F.}\ \bibnamefont {Santos}}, \
  and\ \bibinfo {author} {\bibfnamefont {V.~S.}\ \bibnamefont {Batista}},\
  }\href {\doibase 10.1146/annurev.physchem.040808.090409} {\bibfield
  {journal} {\bibinfo  {journal} {Annual Review of Physical Chemistry}\
  }\textbf {\bibinfo {volume} {60}},\ \bibinfo {pages} {293} (\bibinfo {year}
  {2009})}\BibitemShut {NoStop}%
\bibitem [{\citenamefont {Uhrig}(2008)}]{Uh2008}%
  \BibitemOpen
  \bibfield  {author} {\bibinfo {author} {\bibfnamefont {G.~S.}\ \bibnamefont
  {Uhrig}},\ }\href {http://stacks.iop.org/1367-2630/10/i=8/a=083024}
  {\bibfield  {journal} {\bibinfo  {journal} {New Journal of Physics}\ }\textbf
  {\bibinfo {volume} {10}},\ \bibinfo {pages} {083024} (\bibinfo {year}
  {2008})}\BibitemShut {NoStop}%
\bibitem [{\citenamefont {Morton}\ \emph {et~al.}(2006)\citenamefont {Morton},
  \citenamefont {Tyryshkin}, \citenamefont {Ardavan}, \citenamefont {Benjamin},
  \citenamefont {Porfyrakis}, \citenamefont {Lyon},\ and\ \citenamefont
  {Briggs}}]{Mo2006}%
  \BibitemOpen
  \bibfield  {author} {\bibinfo {author} {\bibfnamefont {J.~J.~L.}\
  \bibnamefont {Morton}}, \bibinfo {author} {\bibfnamefont {A.~M.}\
  \bibnamefont {Tyryshkin}}, \bibinfo {author} {\bibfnamefont {A.}~\bibnamefont
  {Ardavan}}, \bibinfo {author} {\bibfnamefont {S.~C.}\ \bibnamefont
  {Benjamin}}, \bibinfo {author} {\bibfnamefont {K.}~\bibnamefont
  {Porfyrakis}}, \bibinfo {author} {\bibfnamefont {S.~A.}\ \bibnamefont
  {Lyon}}, \ and\ \bibinfo {author} {\bibfnamefont {G.~A.~D.}\ \bibnamefont
  {Briggs}},\ }\href {\doibase 10.1038/nphys192} {\bibfield  {journal}
  {\bibinfo  {journal} {Nat Phys}\ }\textbf {\bibinfo {volume} {2}},\ \bibinfo
  {pages} {40} (\bibinfo {year} {2006})}\BibitemShut {NoStop}%
\bibitem [{\citenamefont {Liu}\ \emph {et~al.}(2013)\citenamefont {Liu},
  \citenamefont {Po}, \citenamefont {Du}, \citenamefont {Liu},\ and\
  \citenamefont {Pan}}]{Liu2013}%
  \BibitemOpen
  \bibfield  {author} {\bibinfo {author} {\bibfnamefont {G.-Q.}\ \bibnamefont
  {Liu}}, \bibinfo {author} {\bibfnamefont {H.~C.}\ \bibnamefont {Po}},
  \bibinfo {author} {\bibfnamefont {J.}~\bibnamefont {Du}}, \bibinfo {author}
  {\bibfnamefont {R.-B.}\ \bibnamefont {Liu}}, \ and\ \bibinfo {author}
  {\bibfnamefont {X.-Y.}\ \bibnamefont {Pan}},\ }\href {\doibase
  10.1038/ncomms3254} {\bibfield  {journal} {\bibinfo  {journal} {Nature
  Communications}\ }\textbf {\bibinfo {volume} {4}},\ \bibinfo {pages} {2254}
  (\bibinfo {year} {2013})}\BibitemShut {NoStop}%
\bibitem [{\citenamefont {L\'opez}\ and\ \citenamefont
  {L\'opez}(2011)}]{LoLo2012a}%
  \BibitemOpen
  \bibfield  {author} {\bibinfo {author} {\bibfnamefont {G.~V.}\ \bibnamefont
  {L\'opez}}\ and\ \bibinfo {author} {\bibfnamefont {P.}~\bibnamefont
  {L\'opez}},\ }\href {\doibase 10.4236/jmp.2012.31013} {\bibfield  {journal}
  {\bibinfo  {journal} {J. Mod. Phys.}\ }\textbf {\bibinfo {volume} {3}},\
  \bibinfo {pages} {85} (\bibinfo {year} {2011})}\BibitemShut {NoStop}%
\bibitem [{\citenamefont {L\'opez}\ and\ \citenamefont
  {L\'opez}(2012)}]{LoLo2012b}%
  \BibitemOpen
  \bibfield  {author} {\bibinfo {author} {\bibfnamefont {P.~C.}\ \bibnamefont
  {L\'opez}}\ and\ \bibinfo {author} {\bibfnamefont {G.~V.}\ \bibnamefont
  {L\'opez}},\ }\href {\doibase 10.4236/jmp.2012.39118} {\bibfield  {journal}
  {\bibinfo  {journal} {J. Mod. Phys.}\ }\textbf {\bibinfo {volume} {3}},\
  \bibinfo {pages} {902} (\bibinfo {year} {2012})}\BibitemShut {NoStop}%
\bibitem [{\citenamefont {Breuer}\ and\ \citenamefont
  {Petruccione}(2002)}]{BrePet02}%
  \BibitemOpen
  \bibfield  {author} {\bibinfo {author} {\bibfnamefont {H.~P.}\ \bibnamefont
  {Breuer}}\ and\ \bibinfo {author} {\bibfnamefont {F.}~\bibnamefont
  {Petruccione}},\ }\href@noop {} {\emph {\bibinfo {title} {The Theory of Open
  Quantum Systems}}}\ (\bibinfo  {publisher} {Oxford University Press, USA},\
  \bibinfo {year} {2002})\BibitemShut {NoStop}%
\bibitem [{\citenamefont {Pasini}\ \emph {et~al.}(2008)\citenamefont {Pasini},
  \citenamefont {Fischer}, \citenamefont {Karbach},\ and\ \citenamefont
  {Uhrig}}]{PaFiKaUh2008}%
  \BibitemOpen
  \bibfield  {author} {\bibinfo {author} {\bibfnamefont {S.}~\bibnamefont
  {Pasini}}, \bibinfo {author} {\bibfnamefont {T.}~\bibnamefont {Fischer}},
  \bibinfo {author} {\bibfnamefont {P.}~\bibnamefont {Karbach}}, \ and\
  \bibinfo {author} {\bibfnamefont {G.~S.}\ \bibnamefont {Uhrig}},\ }\href
  {\doibase 10.1103/PhysRevA.77.032315} {\bibfield  {journal} {\bibinfo
  {journal} {Phys. Rev. A}\ }\textbf {\bibinfo {volume} {77}},\ \bibinfo
  {pages} {032315} (\bibinfo {year} {2008})}\BibitemShut {NoStop}%
\bibitem [{\citenamefont {Plenio}(2005)}]{Plenio2005}%
  \BibitemOpen
  \bibfield  {author} {\bibinfo {author} {\bibfnamefont {M.~B.}\ \bibnamefont
  {Plenio}},\ }\href {\doibase 10.1103/PhysRevLett.95.090503} {\bibfield
  {journal} {\bibinfo  {journal} {Phys. Rev. Lett.}\ }\textbf {\bibinfo
  {volume} {95}},\ \bibinfo {pages} {090503} (\bibinfo {year}
  {2005})}\BibitemShut {NoStop}%
\bibitem [{\citenamefont {Sainz}\ \emph {et~al.}(2011)\citenamefont {Sainz},
  \citenamefont {Burlak},\ and\ \citenamefont {Klimov}}]{Sainz2011}%
  \BibitemOpen
  \bibfield  {author} {\bibinfo {author} {\bibfnamefont {I.}~\bibnamefont
  {Sainz}}, \bibinfo {author} {\bibfnamefont {G.}~\bibnamefont {Burlak}}, \
  and\ \bibinfo {author} {\bibfnamefont {A.~B.}\ \bibnamefont {Klimov}},\
  }\href {\doibase 10.1140/epjd/e2011-20309-7} {\bibfield  {journal} {\bibinfo
  {journal} {The European Physical Journal D}\ }\textbf {\bibinfo {volume}
  {65}},\ \bibinfo {pages} {627} (\bibinfo {year} {2011})}\BibitemShut
  {NoStop}%
\end{thebibliography}%
\end{document}